# Self-Healing Network of Interconnected Edge Devices Empowered by Infrastructure-as-Code and LoRa Communication

Rob Carson, Mohamed Chahine Ghanem * and Feriel Bouakkaz
*Department of Computer Science*, University of Liverpool, Liverpool, UK
*Efrei Research Lab, Paris Pantheon-Assas University,* FRANCE
* Corresponding author:

*Abstract*—This Paper proposes a self-healing, automated network of Raspberry Pi devices designed for deployment in scenarios where traditional networking is unavailable. Leveraging the low-power, long-range capabilities of the LoRa (Long Range) protocol alongside Infrastructure as Code (IaC) methodologies, the research addresses challenges such as limited bandwidth, data collisions, and node failures. Given that LoRa's packet-based system is incompatible with conventional IaC tools like Ansible and Terraform, which rely on TCP/IP networking, the research adapts IaC principles within a containerised architecture deployed across a Raspberry Pi cluster. Evaluation experiments indicate that fragmenting data packets and retransmitting any missed fragments can mitigate LoRa's inherent throughput and packet size limitations, although issues such as collisions and line-of-sight interference persist. An automated failover mechanism was integrated into the architecture, enabling unresponsive services to be redeployed to alternative nodes within one second, demonstrating the system's resilience in maintaining operational continuity despite node or service failures. The paper also identifies practical challenges, including the necessity for time-slotting transmissions to prevent data packet overlap and collisions. Future research should explore the integration of mesh networking to enhance range, develop more advanced scheduling algorithms, and adopt cutting-edge low-power wide-area network (LPWAN) techniques.

## I. INTRODUCTION

Existing network setups in remote environments face significant challenges in providing reliable communication. Key issues include power constraints, geographical barriers (e.g., mountains or dense forests), maintenance difficulties in remote or hostile locations, and susceptibility to equipment destruction in conflict zones. By automating the management and recovery processes using tools such as Ansible and Terraform, the network will ensure resilience and continuous communication even when nodes fail or connections are disrupted. The research seeks to bridge the gap in existing solutions that fail to provide a fully integrated system for edge deployments, where both network resilience and automated infrastructure management are critical [1].

### A. Research Aim and Questions

This research topic looks to develop a self-healing network of interconnected edge devices by exploiting IaC technologies for automated infrastructure management of new and existing edge devices. The edge devices will be connected via LoRa (Long-Range) technology to establish a resilient communication network within hostile/harsh environments. IaC tools are already established as effective through such work as Batu [2], achieving an 85% reduction in recovery time compared to manual intervention, making them a viable option for keeping edge infrastructure operational. LoRa's long-distance and low power consumption make it an obvious candidate for remote environments where more common wireless infrastructure is sparse and has already been concluded as reliable by Borsos et al. [3], when applied to critical infrastructure use cases. Still, work by Fragkopoulos, et al. [4] has highlighted the need for optimised transmission parameters to improve gateway reception in areas with limited coverage [5].

While there are existing implementations of individual components of this thesis proposal, like IaC for automated infrastructure management and LoRa-based communication systems, there is no solution which integrates all discussed technologies to address the challenge of maintaining a resilient network of communicating devices running various applications at the edge. As an example, Batu [2] showed that IaC is better for automation and scalability than traditional backup solutions, but the research is limited to a cloud-based deployment, not an edge deployment. Furthermore, the work conducted by Wagih & Birley [6] is key to highlighting some fundamental ways of improving LoRa connectivity between nodes, but there seems to be no work which tests the limitations of LoRa with specific use cases where applications or services must remain operational. The primary research questions guiding this work are:

**RQ1**: How do IaC tools affect the ease of management of edge device networks?
**RQ2**: How effective are IaC tools in maintaining network availability over LoRa?
**RQ3**: What impact does LoRa communication have on applications/services that must remain operational?
**RQ4**: How does the combination of IaC technologies and LoRa communication enhance network resilience in challenging environments?
These questions guided our adopted research methodology,

test-bed design and experimentation focusing on practical applications and real-world challenges of edge computing.

## II. RELATED WORK

The integration of self-healing capabilities into LoRa-based edge networks addresses critical challenges in reliability, scalability, and autonomy within constrained environments. Recent research has explored various methodologies to enhance network resilience and performance.

### A. LoRa-WAN Connectivity and Antenna Types

Wagih & Birley [6] investigate the impact of antenna choice on LoRa-WAN connectivity, comparing whip and UWB monopole antennas. They find that UWB antennas provide better and more stable RSSI performance, particularly indoors [7]. However, differences were minimal outdoors. This study underscores the necessity of antenna testing based on deployment environments, directly informing the choice of antennas for reliability in diverse scenarios [5].

### B. LoRa in Critical Infrastructure

Borsos et al. [3] explore LoRa's reliability in urban, underwater, and interference-rich environments, demonstrating effective communication in dense urban settings up to 1 km. Crucially, higher spreading factors provided better resistance to interference but at the cost of reduced data rates. The authors advocate adaptive data rate (ADR) mechanisms to dynamically manage these trade-offs. Their insights are particularly relevant for critical and military infrastructure subject to cyber-attacks [8].

### C. LoRaWAN Performance Factors

Fragkopoulos et al. [4] assess LoRaWAN performance optimisation through parameters like acknowledgement mechanisms, gateway availability, and spreading factor allocation. Findings show gateway congestion with acknowledgement mechanisms despite improved packet delivery ratios. They recommend distributing gateways and adaptive allocation of transmission power based on node distances to enhance reliability and efficiency, especially relevant in power-constrained edge deployments [9].

### D. LoRaMesher for Mesh Networks

Solé et al. [10] detail the LoRa Mesher library, enabling adaptive mesh topology for LoRa nodes. This approach enhances network resilience through dynamic routing, reliable message forwarding, and decreased dependence on gateways. The mesh topology addresses the traditional range limitations of LoRa networks, relevant to urban deployments and scenarios requiring high resilience.

### E. Kubernetes with Raspberry Pi

Füstös et al. [11] evaluate Kubernetes clusters on Raspberry Pis for edge applications, utilising Ansible for automation. Practical demonstrations confirm feasibility for diverse workloads, highlighting essential elements like Prometheus monitoring and Grafana visualisation. However, scalability and additional security measures, especially against physical access, remain areas requiring further exploration [12].

### F. IaC and Disaster Recovery

Batu [2] investigates IaC's effectiveness for disaster recovery in cloud environments, achieving an 85% reduction in recovery time with Terraform. Though results are promising for automated resilience, its applicability to wireless node clusters remains uncertain due to differing connectivity and physical deployment challenges, indicating the need for further research tailored specifically for edge scenarios.

### G. Multi-Hop LoRa Networks

Wong et al. [13] discuss extending LoRa's range and reliability through multi-hop networks using reactive routing protocols like AODV, achieving high packet delivery ratios. Security concerns due to multiple transmission points require encryption and mutual authentication strategies. Multi-hop communication effectively addresses single-hop LoRa network limitations, particularly beneficial for remote deployments.

### H. LoRa Communication in Challenging Environments

Maleki et al. [14] highlight LoRa's robustness using CSS modulation, offering interference resistance crucial for IoT deployments in challenging environments. Povalac et al. [15] reveal practical synchronisation issues and encryption vulnerabilities through real-world traffic analysis, emphasising the need for improved network resilience strategies and rigorous security measures.

### I. Failure Recovery for IoT Applications

Olorunnife et al. [16] propose container-based failure recovery for remote IoT applications, achieving automatic redeployment and reduced downtime. While effective, increased container sizes significantly impact recovery times. The study provides foundational insights for this thesis's self-healing network objective, suggesting smaller, readily available backup containers to mitigate recovery delays [17] and [18].

To sum up, despite these advancements, challenges remain in implementing comprehensive self-healing mechanisms in LoRa-based edge networks. Issues such as limited bandwidth, duty cycle restrictions, and the need for lightweight protocols necessitate innovative solutions that balance performance with resource constraints. This paper contributes to the field by presenting a modular, containerised architecture that facilitates autonomous fault detection and recovery, tailored to the specific limitations and requirements of LoRa-based edge environments.

## III. RESEARCH METHODOLOGY

### A. Experimental Setup

This research creates a network of five Raspberry Pi devices which all communicate via LoRa. Each Pi has a "management" container which executes several tasks, including sending and receiving file updates, collecting local metrics, and managing failover operations if a Pi goes offline or fails [19].

One Pi also runs an InfluxDB container to store all metrics, while another Pi also runs a Grafana container to visualise these metrics in real time. A high-level architecture diagram of the IT artefact can be seen in Figure 1. The hardware making up this research is five Raspberry Pi 4 boards, each linked to an Adafruit LoRa Bonnet (915 MHz) and an 868 MHz LoRa antenna. The use of Raspberry Pi boards means that the system has enough processing power to manage the necessary containers while remaining compact and energy-efficient.

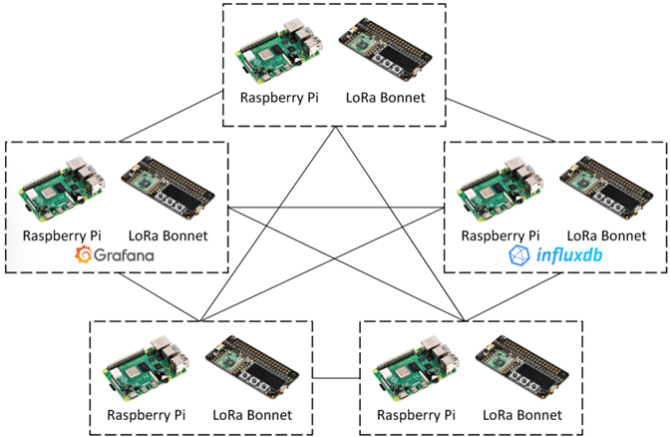

**Fig. 1:** *High-Level Pi Network Architecture*

### B. Key Design Features

- **Containerisation and Failover**: Uses Docker containers for consistent and reproducible deployment. A heartbeat mechanism via LoRa identifies node failures, automatically initiating failover by relocating essential services to operational nodes.
- **LoRa Communication**: Manages low bandwidth and high latency constraints by fragmenting larger payloads for efficient transmission and reassembly.
- **Metrics and Testing**: Pis independently record local metrics like CPU usage and LoRa packet latency, which are aggregated in InfluxDB. Testing simulates node failures and varying LoRa parameters to evaluate system resilience.

### C. Software Modules

The orchestration layer is defined in *docker-compose.yml*, which mounts the SPI/GPIO devices required by the RFM9x radio, exposes `/var/run/docker.sock` so the management service can manipulate local containers, and connects every service to an internal Docker network. Each node builds its management image from a minimal Python 3.9 base (Dockerfile) and installs only the libraries needed for LoRa access, GPIO control, and Docker interaction, ensuring a reproducible footprint across the cluster.

Runtime configuration resides in *host_utils.py*, which centralises radio parameters, heartbeat intervals, and the desired service layout while exposing `get_real_hostname()` for deterministic node identification. Low-level radio access is wrapped by *lora_radio.py*, providing send/receive primitives with CRC checking. Above this layer, *reliable_sender.py* and *reliable_receiver.py* fragment application payloads (<252 bytes per LoRa frame), base64-encode them, and reassemble them with selective retransmission; a thread-level lock serialises radio access to prevent collisions. System health is captured by *metrics_collector.py*, which samples CPU, memory, and container status via *psutil* and the Docker SDK, timestamping each packet so end-to-end latency can be derived. On the designated Influx host, *influx_ingestor.py* writes reassembled metrics to the database; on all other nodes, it simply forwards packets. Code and configuration consistency is maintained by *git_sync_manager.py*: the script watches specified directories, commits edits locally, packs them into Git bundles, and transmits those bundles via LoRa so every node can self-update. High availability is enforced by *failover_manager.py*, which monitors heartbeats, declares nodes offline after a timeout, and launches replacement services with the Docker SDK following the intended layout. Finally, *main.py* initialises the radio and fail-over tables, then spawns the sender, receiver, metrics, and Git threads. Its behaviour adapts automatically to the node's role—InfluxDB server, Grafana visualiser, or ordinary edge node—ensuring autonomous operation and rapid recovery across the LoRa-based cluster.

## IV. TESTING AND EVALUATION RESULTS

All measurements were taken in a semi-urban domestic setting with the five Pis distributed across two floors of a brick building. Unless otherwise stated, the radios operated at 868MHz with 20dBm output power, spreading factor 7, coding rate 4/5, and 125kHz bandwidth—the UK 1% duty-cycle limit was respected throughout. Latency—the end-to-end delay between metric timestamp and successful ingestion in InfluxDB—served as the primary performance indicator.

### A. Node Count Stability

Figure 2 shows that up to four transmitters maintain a tight 50–60s latency band, punctuated only by isolated peaks where fragment collisions trigger selective retransmission.

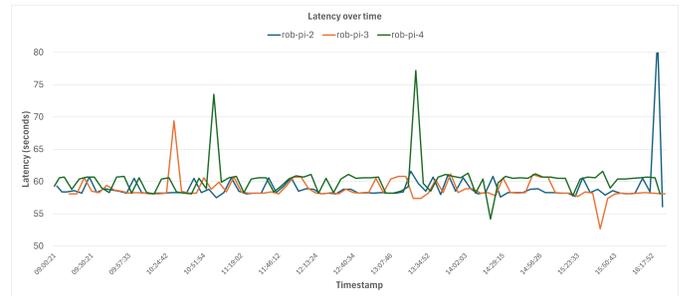

**Fig. 2:** Latency with 4 nodes

### B. Reporting Interval and Synchronisation Effects

Doubling the metric interval to ten minutes, illustrated in Figure 3 restores headroom: both spike frequency and magnitude fall, confirming that contention, not intrinsic radio delay, is the bottleneck.

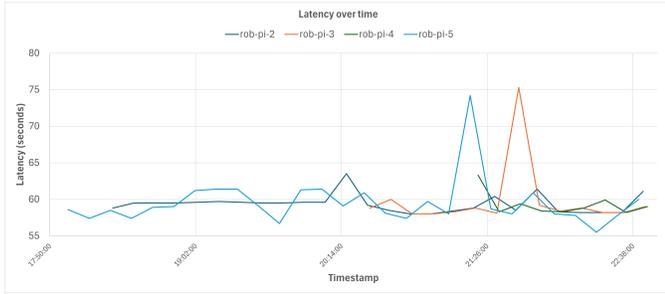

**Fig. 3:** Latency with 10-minute interval

## C. Varying LoRa Settings

To evaluate how LoRa settings influence latency, three key parameters were swept—bandwidth, coding rate, and transmission power—while the spreading factor remained at its default value due to known Adafruit-library issues. With the bandwidth widened to 500kHz (Figure 4), most latencies fell into the 20–30s band, though occasional 40–90s spikes still signalled brief collisions or congestion: higher throughput cannot completely mask traffic bursts. Raising the coding rate to 4/8 (Figure 5) shifted typical performance to 45–70s with few peaks, a predictable trade-off where extra redundancy shields against corruption at the expense of airtime. Dropping transmit power to 5dBm (Figure 6) left latencies largely unchanged (40–60s with sporadic 90–100s spikes) because the Pis sit close together; over wider deployments, the same reduction would conserve energy at the cost of range.

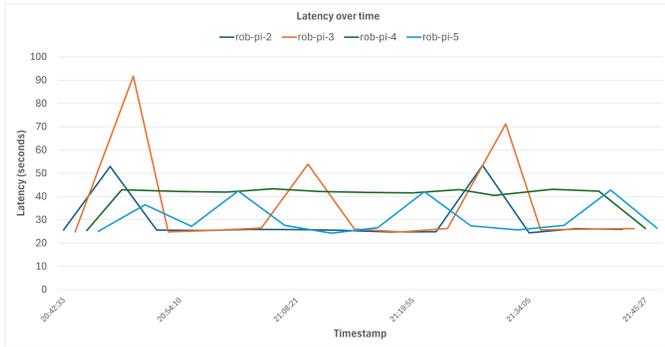

**Fig. 4:** Latency with bandwidth of 500 kHz

## D. Service Failover Robustness

Fail-over trials with Docker images of 8.8MB, 339MB, and 5.5GB, as summarised in Table I, are all completed in 1s. Because every node maintains an up-to-date image cache, downtime is dominated by container start-up rather than image transfer. This validates the design goal of infrastructure self-healing irrespective of service weight, a critical feature for unattended deployments where manual intervention is impossible.

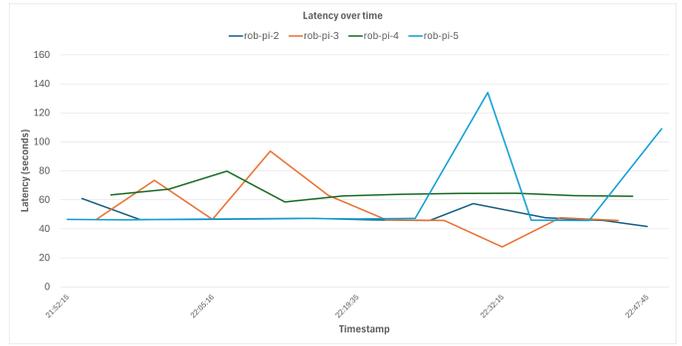

**Fig. 5:** Latency with coding rate of 4/8

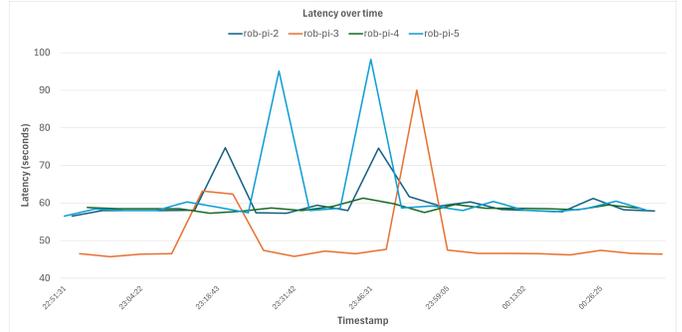

**Fig. 6:** Latency with transmission power of 5 dBm

## V. RESULTS AND DISCUSSION

The test cases showcased in this chapter give an overview of how the IT artefact performed under varying conditions. One noticeable finding was the system's latency sensitivity as the number of nodes increased, with the frequency of packet collisions rising with an increased number of nodes, leading to spikes in delivery times and some data loss. This highlights one of LoRa's main drawbacks. While it does have long-range communication, its lack of built-in collision avoidance becomes an issue when nodes begin overlapping when sending data. This emphasises the need to incorporate strict scheduling techniques or an adaptive data rate to ensure collisions are minimised. When tests investigated varying metric intervals, it was clear that increasing the interval allowed nodes more time to transmit and re-transmit all data fragments with fewer instances of overlap. Conversely, starting all nodes at the same time created much higher latencies and periods of prolonged data loss, reinforcing the need for scheduling data transmissions to avoid data loss and latency spikes.

Despite overlap challenges, the fragmentation and fragment retransmission feature was very effective in preventing data from being missed, even for the largest payloads. Testing latency with varying Git bundle file sizes did not produce an exponential relationship, which is a key strength of this feature, as it showcases that there is no limit to the scalability of synchronising files or sending data. This also enables large packets of data to be sent reliably without the risk of corruption. Failover tests with containerised services of drastically different image sizes highlighted the consistency of this feature, where services were redeployed in approximately one second every time. This is mainly due to each node already having the latest image stored locally, but still showing no constraints on the failover system with container sizes. This further validates one of the key aspects of this research, of creating a self-healing network in remote environments – achieving instant redeployment of services, no matter their size, highlighting this functionality.

**TABLE I:** Fail over time with varying image size

| Image Size (MB) | Time (seconds)l | | | | | | |
|---|---|---|---|---|---|---|---|
| 8.83 | 1.33 | 1.31 | 1.02 | 0.98 | 0.93 | 0.92 | 0.85 |
| 339 | 0.94 | 1.53 | 0.98 | 1.28 | 1 | 1.33 | 0.92 |
| 5470 | 1.17 | 1.07 | 1.3 | 1 | 1.31 | 0.9 | 1 |

## VI. CONCLUSION

This work closes a gap by combining LoRa communication with container-based self-healing—prior studies treated these aspects separately. For industry, it offers a practical template for low-cost, long-range monitoring where conventional connectivity is unavailable, though throughput limits rule out data-intensive workloads. Some recommendations for taking this work further are as follows: integrate adaptive scheduling to keep nodes within duty-cycle limits; explore mesh or multi-hop routing to bypass obstacles; field-test weather-proofed enclosures; enrich fail-over decisions with live resource metrics; and benchmark LoRa against alternative LPWANs such as Sigfox, NB-IoT, or ZigBee. Security is also an area requiring attention; for example, lightweight encryption and secret management are mandatory if the system operates autonomously in hostile environments. Finally, this research highlighted that traditional IaC tools hindered management over LoRa, but IaC principles re-implemented in containers maintained availability. LoRa's constrained link demands careful scheduling and fragmentation, yet, coupled with containerised fail-over, it can still underpin a resilient edge network in challenging environments.